\documentclass[prl,twocolumn,showpacs,superscriptaddress,preprintnumbers]{revtex4-2}
\usepackage{amsmath,amssymb,amsfonts,float,graphics,epsfig,epstopdf,color,verbatim,tabularx,bm,multirow,appendix}
\usepackage{tikz}
\usepackage{amsmath,amssymb,graphicx}
\usepackage{wasysym}
\usepackage[utf8]{inputenc}
\usepackage[T1]{fontenc}
\usepackage{xcolor}
\usepackage{dsfont}
\usepackage{tabularx}
\usepackage{array}

\usepackage{textcomp}
\usepackage{bm}

\usepackage{graphicx}%
\usepackage{xcolor}
\usepackage{dcolumn}
\usepackage{bm}
\usepackage{xspace}
\usepackage{time}
\usepackage{booktabs}
\usepackage{multirow}
\usepackage{makecell}

\pdfoutput=1
\usepackage{color}
\definecolor{LinkColor}{rgb}{0.256,0.439,0.588}
\usepackage{hyperref}
\hypersetup{
colorlinks=true,
citecolor=LinkColor,
linkcolor=LinkColor,
urlcolor=LinkColor
}

\def\bk{{\mathbf{k}}}

\def\bq{{\mathbf{q}}}
\def\bQ{{\mathbf{Q}}}
\def\bG{{\mathbf{G}}}

\newcommand{\newsect}[1]{\noindent \textit{\textcolor{blue}{#1.-}}}

\begin{document}
\title{Momentum-space quantum Monte Carlo for twisted bilayer graphene: \\ Interaction effects in the presence of remote bands}

\author{Shibo Shan}
\affiliation{Department of Physics and HK Institute of Quantum Science \& Technology, The University of Hong Kong, Pokfulam Road,  Hong Kong SAR, China}
\affiliation{State Key Laboratory of Optical Quantum Materials, The University of Hong Kong, Pokfulam Road,  Hong Kong SAR, China}

\author{Cheng Huang}
\affiliation{Department of Physics and HK Institute of Quantum Science \& Technology, The University of Hong Kong, Pokfulam Road,  Hong Kong SAR, China}
\affiliation{State Key Laboratory of Optical Quantum Materials, The University of Hong Kong, Pokfulam Road,  Hong Kong SAR, China}


\author{Zi Yang Meng}
\email{zymeng@hku.hk}
\affiliation{Department of Physics and HK Institute of Quantum Science \& Technology, The University of Hong Kong, Pokfulam Road,  Hong Kong SAR, China}
\affiliation{State Key Laboratory of Optical Quantum Materials, The University of Hong Kong, Pokfulam Road,  Hong Kong SAR, China}

\date{March 2026}

\date{March 2026}

\begin{abstract}
Twisted bilayer graphene hosts strong correlation effects within topological flat bands separated from remote dispersive bands. This separation motivates projection onto the central bands, which greatly reduces the computational complexity of many-body calculations, including momentum-space quantum Monte Carlo (QMC) simulations~\cite{zhangMomentum2021}. However, the interaction-induced Mott gap obtained in the projected model can be comparable to the noninteracting flat--remote band separation, raising the question of whether the projection remains valid. Here, we extend momentum-space QMC to explicitly include remote BM bands within a broad energy window at charge neutrality. We find that the projected description remains reliable over a broad parameter regime, up to $u_0/u_1\lesssim0.9$: the projected and remote-band calculations yield similar low-energy electron and trion spectra, and the Kramers intervalley-coherent ground state remains stable. In particular, the Dirac trion~\cite{shanTrion2026} remains robust in the finite-temperature normal state when remote bands are included. Near the band-touching limit $u_0=u_1$, remote-band electrons acquire substantial low-energy spectral weight, accompanied by suppression of the trion weight and loss of the Kramers intervalley-coherent state. Our findings establish the broad validity of flat-band projection while reconciling the projected and THF viewpoints by identifying the near-band-touching regime in which explicit remote-band degrees of freedom become essential.
\end{abstract}

\maketitle

\newsect{Introduction} 
Twisted bilayer graphene (TBG) has emerged as a highly tunable platform for studying the interplay of strong correlations, topology, and quantum geometry. Near the magic angle, the suppression of electronic kinetic energy produces a rich phase diagram of superconducting and correlated insulating states that can be tuned by carrier density and external fields~\cite{caoCorrelated2018,cao2018unconventional,lu2019superconductors,stepanov2020untying,liu2021tuning,saito2020independent,arora2020superconductivity,yankowitz2019tuning,ohEvidenceUnconventionalSuperconductivity2021,tanaka2025superfluid,parkExperimental2026,hyunjinResoving2026}. These insulating phases include states with broken spin, valley, or lattice symmetries, orbital magnetism, and nontrivial topology, including integer and fractional Chern insulators~\cite{sharpe2019emergent,kerelskyMaximized2019,choiElectronic2019,jiangCharge2019,nuckolls2020strongly,serlin2020intrinsic,xie2021fractional,saito2021hofstadter,choi2021correlation,wu2021chern,pierce2021unconventional,KwanKekule,nuckolls2023quantum}. Thermodynamic and spectroscopic measurements have further revealed cascades of flavor-symmetry-breaking transitions, large flavor entropy, local-moment behavior, strong mass renormalization, and pronounced interaction-driven reconstruction of the electronic spectrum~\cite{wongCascade2020,zondinerCascade2020,rozen2021Entropic,saitoIsospinPomeranchukEffect2021,zhangHeavy2025}. More recently, the quantum twisting microscope (QTM) has enabled momentum-resolved imaging of the interaction-renormalized moiré bands~\cite{inbarQuantum2023,xiao2026imaging}. Together, these observations demonstrate that the low-energy physics of TBG is governed by a delicate interplay of band structure, topology, and Coulomb interactions, raising the fundamental question of which electronic bands must be retained in an effective many-body description.



\begin{figure}[htp!]
\centering
\includegraphics[width=\columnwidth]{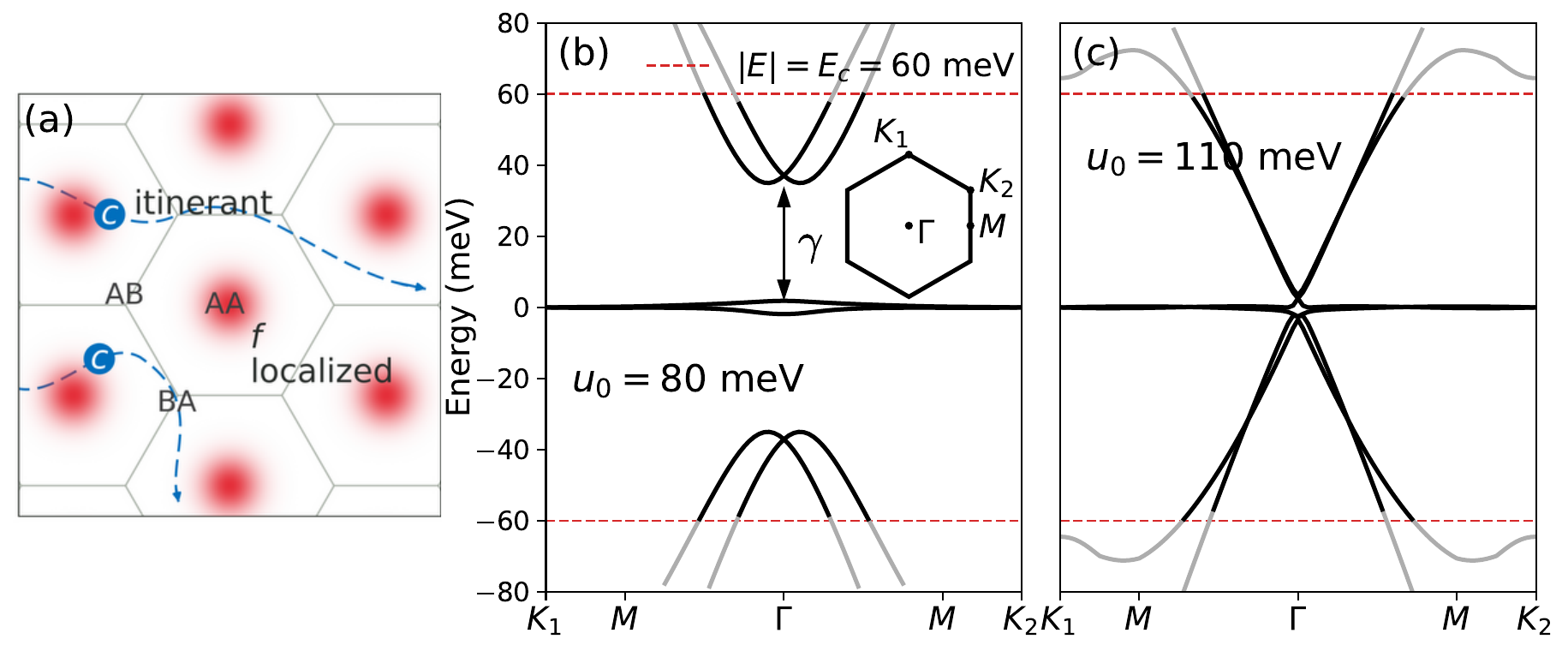}
\caption{\textbf{Real-space $f$ and $c$ electrons schematic and their dispersions in the bare BM model.} (a) Schematic of the topological heavy-fermion description, showing one localized $f$ orbital (red) per moiré unit cell and itinerant $c$ electrons (blue) propagating across the lattice. (b,c) Bare BM dispersions along $K_1\!-\!M\!-\!\Gamma\!-\!M\!-\!K_2$ at $\theta=1.08^\circ$ and $u_1=110\,\mathrm{meV}$, with $u_0=80\,\mathrm{meV}$ in (b) and $u_0=110\,\mathrm{meV}$ in (c). The arrow in (b) denotes the flat--remote band separation $\gamma$ at $\Gamma$, which closes at the band-touching point $u_0=u_1$ in (c). Black and gray segments denote states inside and outside the energy window $\lvert E\rvert\leq E_c=60\,\mathrm{meV}$, respectively. The inset in (b) shows the moiré Brillouin zone and its high-symmetry points.}
\label{fig:fig1}
\end{figure}

Theoretically, the Bistritzer--MacDonald (BM) continuum model~\cite{tramblyLocalization2010,bistritzerMoire2011} is frequently used to model the electron kinetics of TBG~\cite{zhangMomentum2021,bernevig2021TBGI,hofmann2022fermionic}. In the dispersion of BM model, the two central flat bands are separated from the nearest remote bands by a direct single-particle gap $\gamma$ at the $\Gamma$ point. At $\theta=1.08^\circ$ and $u_0=80~\mathrm{meV}$, this separation is approximately $\gamma\sim40~\mathrm{meV}$ [Fig.~\ref{fig:fig1}(b)], providing the usual justification for projecting the interacting problem onto the two central bands. However, in momentum space, a screened Coulomb interaction with a characteristic scale of $\sim10~\mathrm{meV}$ can generate a Mott gap of approximately $20~\mathrm{meV}$ at charge neutrality in the projected model~\cite{bultinckGround2020,zhangMomentum2021,hofmann2022fermionic,song2022magic,huangAngle2025}. The resulting many-body gap is therefore no longer parametrically smaller than $\gamma$. Furthermore, increasing $u_0$ reduces $\gamma$ until it vanishes at the band-touching point $u_0=u_1$ [Fig.~\ref{fig:fig1}(c)]. The loss of this separation of energy scales raises a central question: over what parameter range is the two-band projection reliable, and when must the remote bands be retained as active low-energy degrees of freedom?

The topological heavy-fermion (THF) model of TBG~\cite{song2022magic} provides an alternative description of the low-energy BM bands in terms of localized $f$ orbitals centered at the AA regions and itinerant $c$ electrons carrying the band topology, as illustrated in Fig.~\ref{fig:fig1}(a). In momentum space, the central bands are predominantly $f$-like over most of the mBZ, while the hybridization between the $f$ and $c$ sectors becomes important near the $\Gamma$ point [Fig.~\ref{fig:fig1}(b,c)]. The itinerant $c$ sector supplies the additional degrees of freedom required to reproduce the symmetry representations and topology of the BM bands. Within this framework, the large interaction-induced gap is naturally associated with the strong local repulsion between the $f$ electrons~\cite{song2022magic,shi2022heavy,hu2023kondo,hu2023symmetric}.


The THF model provides a unified description of real-space localization and momentum-space topology and has been studied analytically~\cite{cualuguaru2023twisted,hu2023kondo}, using mean-field theories~\cite{song2022magic,shi2022heavy,chou2023kondo,rai2024dynamical,wang2025electron}, and within dynamical mean-field theory~\cite{rai2024dynamical,zhou2024Kondo,hu2023symmetric}. A direct and large-scale QMC treatment of the THF model, however, remains unavailable. Against this background, a central physical question is whether projection onto the central bands is sufficient to capture the low-energy physics of interacting TBG or whether the remote BM bands must be retained explicitly. This question is particularly important because projected QMC studies have already revealed insulating phases, symmetry-breaking orders, thermodynamic behavior, and excitation spectra at integer fillings~\cite{zhangMomentum2021,hofmann2022fermionic,panDynamical2022,panThermodynamic2023,huangEvolution2024,ledwithNonlocal2024,shanTrion2026}, while projected QMC and exact diagonalization have also captured strain effects consistent with experiments~\cite{huang2026strain,hofmann2026monte}. Determining whether these results remain valid after the remote bands are included therefore provides a direct test of the range of validity of the projected description.

More recently, calculations within the projected model identified the Dirac trion---an emergent composite fermionic excitation orthogonal to the microscopic electron operator---as a dominant low-energy charged excitation near the $\Gamma$ point in the topological Mott regime~\cite{ledwithNonlocal2024,ledwithExotic2025,shanTrion2026}. Rather than originating from a conventional Mott state with frozen local moments, this excitation reflects mixed-valence Mott physics in which local charge fluctuations remain active~\cite{zhao2025mixed,zhao2025ancillatheorytwistedbilayer,zhangEmergent2026,danuZooming2021}. The resulting spectrum is strongly momentum differentiated, with a large Mott gap over most of the mBZ but low-energy composite excitations near $\Gamma$~\cite{ledwithNonlocal2024,ledwithExotic2025}. Whether the Dirac trion and other results obtained from momentum-space QMC for the projected Hamiltonian~\cite{zhangMomentum2021,panDynamical2022,hofmann2022fermionic,panThermodynamic2023,huangEvolution2024,huangAngle2025,huang2026strain,shanTrion2026} remain intact when remote bands are explicitly included is therefore an important open question. Addressing this question connects the projected and heavy-fermion limits of TBG.

In this work, we address this question by explicitly retaining the remote BM bands within a broad energy window in continuous-field momentum-space QMC simulations~\cite{zhangMomentum2021,hofmann2022fermionic,zhangFermion2022,huangEvolution2024,huangAngle2025,huang2026strain} at charge neutrality. We calculate the ground-state intervalley-coherent order parameter and the finite-temperature excitation spectra of electrons and Dirac trions~\cite{ledwithExotic2025,shanTrion2026}. For \(u_0\lesssim0.9u_1\), the projected and remote-band calculations yield similar low-energy spectra over several twist angles, demonstrating the broad validity of flat-band projection and the robustness of Dirac-trion physics against remote-band mixing; the intervalley-coherent insulating state also remains stable in the same parameter range. As \(u_0\) approaches the band-touching point \(u_0=u_1\), the flat--remote band separation \(\gamma \to 0\), and remote-band electrons acquire substantial zero-energy spectral weight at both ground state and finite temperatures, with a sharper onset in the former and a smoother crossover in the latter, accompanied by gradual suppression of the trion weight. These results establish the broad validity of flat-band projection while revealing its breakdown near band touching, where itinerant remote-band electrons participate directly in the low-energy excitations and their explicit inclusion, as in the THF description, becomes necessary.

\begin{figure*}[htp!]
    \centering
    \includegraphics[width=0.99\textwidth]{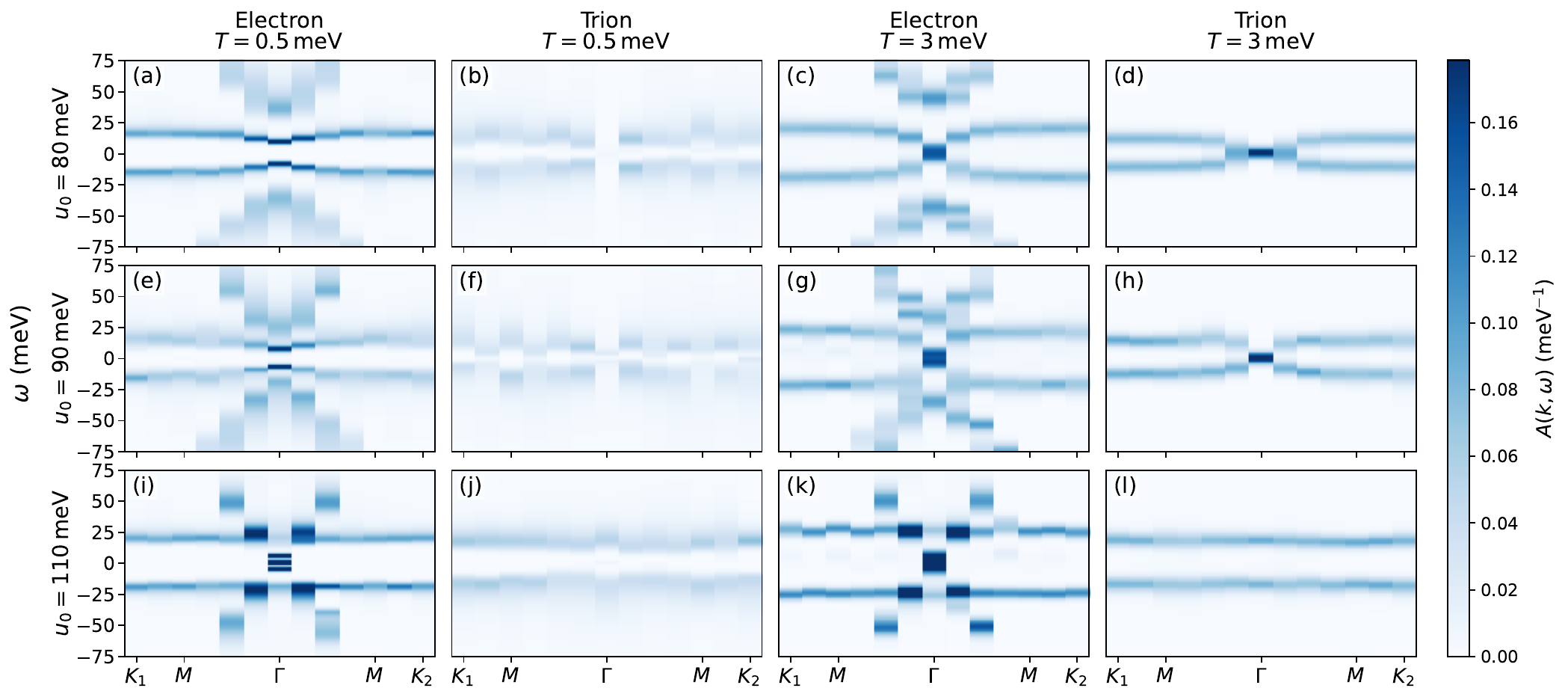}
    \caption{\textbf{Electron and trion spectra with remote bands.}
    QMC obtained electron spectral function $A_e(k,\omega)$ and trion spectral function $A_F(k,\omega)$ along the $K_1$--$M$--$\Gamma$--$M$--$K_2$ path at $\theta=1.08^\circ$. From top to bottom, the rows correspond to $u_0=80\,\mathrm{meV}$, $90\,\mathrm{meV}$, and $110\,\mathrm{meV}$, respectively. From left to right, the columns show the electron and trion spectra at $T=0.5$ meV, followed by the corresponding spectra at $T=3$ meV. All calculations use $L=9$. The energy cutoff is $E_c=80\,\mathrm{meV}$ for $u_0=80\,\mathrm{meV}$ and $90\,\mathrm{meV}$, and $E_c=60\,\mathrm{meV}$ for $u_0=110\,\mathrm{meV}$.}
    \label{fig:fig2}
\end{figure*}

\newsect{Model and Observables} The single-particle Hamiltonian is taken from the BM continuum model~\cite{tramblyLocalization2010,bistritzerMoire2011,huangAngle2025}. We fix the intersublattice interlayer hopping to $u_1=110~\mathrm{meV}$ and vary the intrasublattice interlayer hopping $u_0$. Projecting the Coulomb interaction onto a finite set of BM bands gives~\cite{huangAngle2025,huang2026strain}
\begin{equation}
H=
\sum_{\bk n s \eta}
h^{\mathrm{BM}}_{n\eta}(\bk)
c_{\bk ns\eta}^{\dagger}c_{\bk ns\eta}
+
\frac{1}{2\Omega}
\sum_{\bQ}
V(\bQ)\,
\delta\rho_{-\bQ}\delta\rho_{\bQ},
\label{eq:eq1}
\end{equation}
where $\Omega$ is the sample area and $h^{\mathrm{BM}}_{n\eta}(\bk)$ denotes the BM band energy measured from charge neutrality. The indices $\bk$, $n$, $s=\uparrow,\downarrow$, and $\eta=\pm$ label the mBZ momentum, BM band, spin, and valley, respectively. At each momentum, we retain all BM bands satisfying $\lvert h^{\mathrm{BM}}_{n\eta}(\bk)\rvert\leq E_c$, as illustrated in Fig.~\ref{fig:fig1}(b,c), so the number of retained bands can depend on momentum. We refer to the two central bands as the $C$ sector and all additional retained remote bands as the $R$ sector. The operator $\delta\rho_{\bQ}$ is the density fluctuation projected onto the retained BM bands, whose explicit form is given in the End Matter. We employ the single-gate-screened Coulomb interaction $V(\bQ)=e^2\left(1-\mathrm{e}^{-\lvert\bQ\rvert d}\right)/(2\epsilon\lvert\bQ\rvert)$, with dielectric permittivity $\epsilon=7\epsilon_0$ and gate distance $d/2=20~\mathrm{nm}$~\cite{liuNematic2021}. Following Refs.~\cite{panThermodynamic2023,panDynamical2022}, the momentum transfer $\bQ$ is restricted by a cutoff of the order of the moiré reciprocal-lattice scale.
 
Following Ref.~\cite{ledwithExotic2025,shanTrion2026}, we define the trion operator as 
\begin{equation}
F_{\mathbf{k}\sigma s\eta}
= \frac{1}{\sqrt{N}}
\sum_{\mathbf{R}}
e^{-i\mathbf{k}\cdot\mathbf{R}}
\bigl\{ c_{\mathbf{R}\sigma s\eta},\, \delta n_{\mathbf{R}} \bigr\},
\label{eq:eq2}
\end{equation}
where \(c_{\mathbf R\sigma s\eta}
=
\frac{1}{\sqrt N}
\sum_{\mathbf k\in{\rm mBZ}}
e^{i\mathbf k\cdot\mathbf R}
c_{\mathbf k\sigma s\eta}\) and \(\delta n_{\mathbf R}
=
\sum_{\sigma,s,\eta}
c_{\mathbf R\sigma s\eta}^{\dagger}
c_{\mathbf R\sigma s\eta}
-4 .\)
Here $\mathbf{R}$ labels moir\'e unit cells and \(\sigma=\pm\) labels the sublattice-polarized (Chern) basis within the two projected BM flat bands~\cite{bultinckGround2020,ledwithreview2021}. See Ref.~\cite{shanTrion2026} for the details of this basis construction and End Matter for the QMC measurement of the electron and trion Green's functions and spectral functions.

To characterize the Kramers intervalley-coherent (KIVC) order at charge neutrality~\cite{bultinckGround2020,liaoCorrelation2021}, we define the order parameter as $O_{\mathrm{KIVC}}(\bq)=\sum_{\bk,s}c_{\bk+\bq,s}^{\dagger}\mathcal{M}c_{\bk,s}$, with the ordering wave vector $\bq=\Gamma$. In the plane-wave basis, $\mathcal{M}=\eta_x\mu_y\sigma_x$, where $\eta_x$, $\mu_y$, and $\sigma_x$ act in the valley, layer, and sublattice spaces, respectively. The corresponding structure factor is $S_{\mathrm{KIVC}}(\bq)=\langle O_{\mathrm{KIVC}}^{\dagger}(\bq)O_{\mathrm{KIVC}}(\bq)\rangle/N$, where $N=L^2$ is the number of momenta in the discretized mBZ.

We solve the Hamiltonian in Eq.~\eqref{eq:eq1} and measure the observables defined above using the continuous-field momentum-space QMC method~\cite{zhangMomentum2021,huangAngle2025,huang2026strain,shanTrion2026}. Further computational details are provided in  Ref.~\cite{shanTrion2026}.

\begin{figure}[htp!]
    \centering
    \includegraphics[width=\columnwidth]{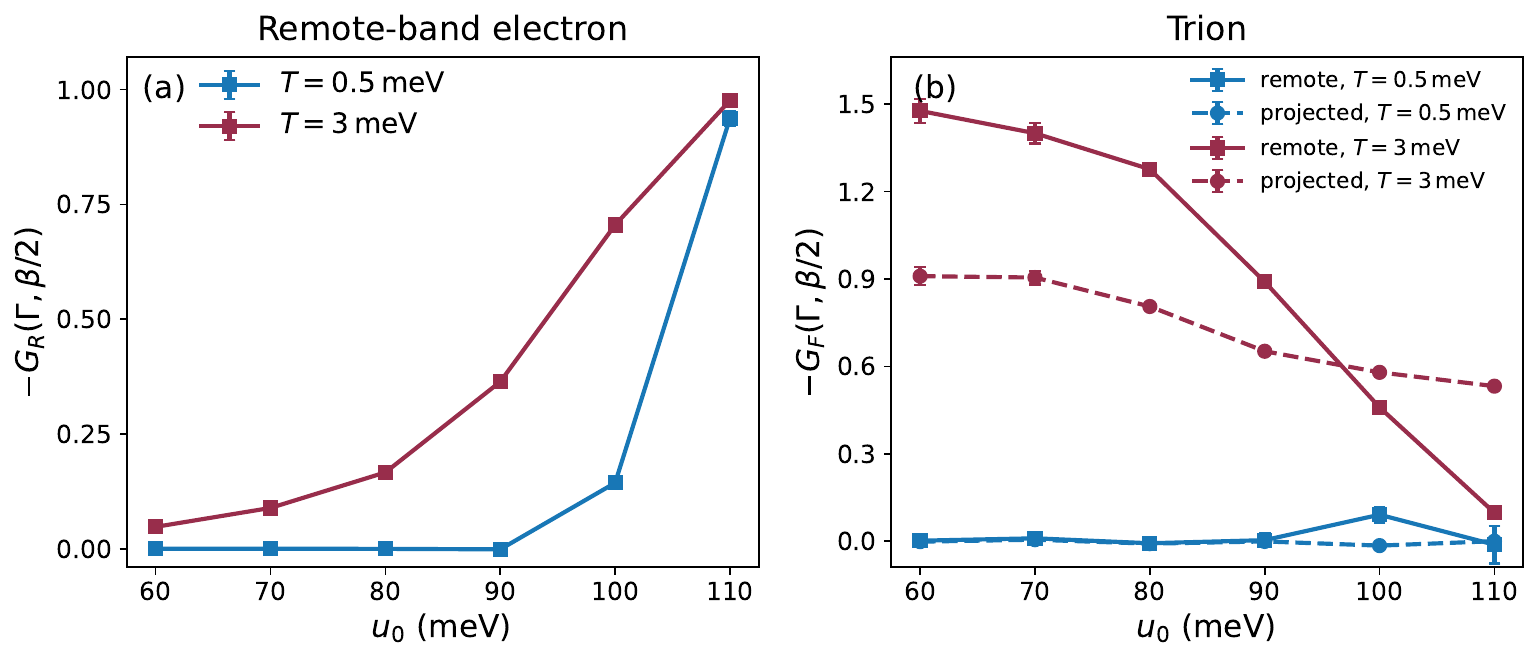}
    \caption{\textbf{Evolution of low-energy remote-band electron and trion spectral weights.}
    Imaginary-time midpoint Green's functions at the $\Gamma$ point, which serve as proxies for the spectral weight near zero energy, are shown as functions of $u_0$ at $\theta=1.08^\circ$. (a) Summed remote-band electron Green's function, $-G_R(\Gamma,\beta/2)=-\sum_{n\in R}G_n(\Gamma,\beta/2)$, where $R$ denotes all retained remote bands outside the two central bands. (b) Trion Green's function, $-G_F(\Gamma,\beta/2)$, for calculations including remote bands and for two-band projected calculations. Solid lines with square markers denote calculations including remote bands with the energy cutoff $E_c=80~\mathrm{meV}$, whereas dashed lines with circular markers denote two-band projected calculations. Blue and red curves correspond to $T=0.5~\mathrm{meV}$ and $T=3~\mathrm{meV}$, respectively.}
    \label{fig:fig3}
\end{figure}

\begin{figure}[htp!]
    \centering
    \includegraphics[width=\columnwidth]{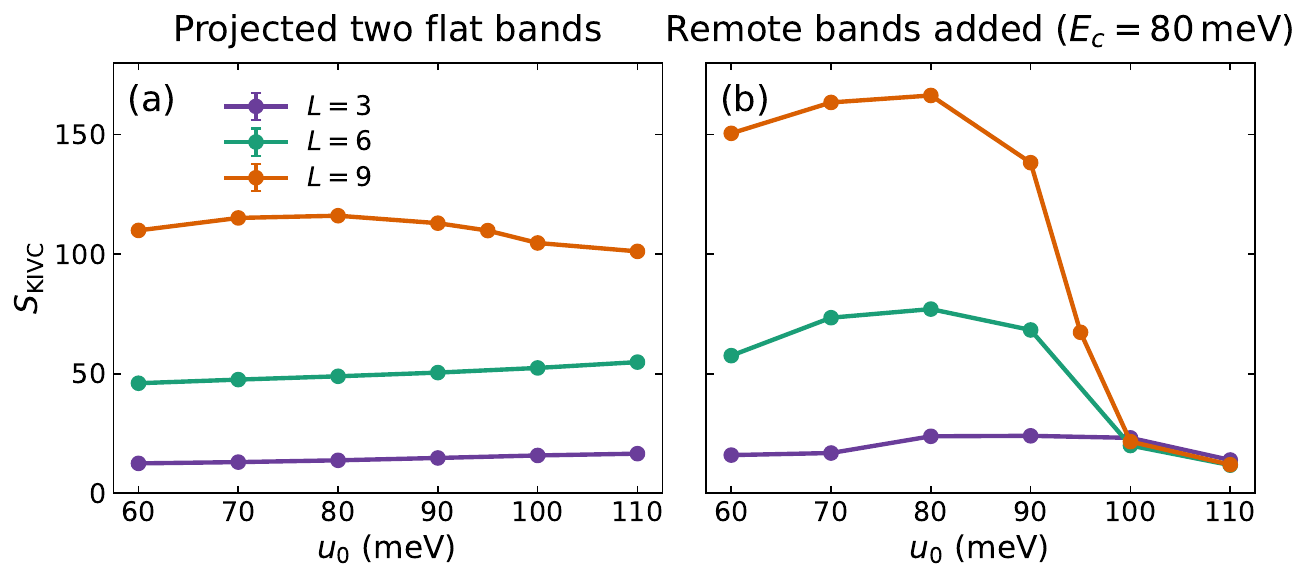}
    \caption{\textbf{Finite-size behavior of KIVC correlations with and without remote bands.}
    The KIVC structure factor
    $S_{\mathrm{KIVC}}
    =\langle O_{\mathrm{KIVC}}^{\dagger}(\Gamma)
    O_{\mathrm{KIVC}}(\Gamma)\rangle/N$
    is shown as a function of $u_0$ at $T=0.5\,\mathrm{meV}$ and
    $\theta=1.08^\circ$, where $N=L^2$ is the number of moiré momenta.
    (a) Results in the projected two-flat-band model.
    (b) Results including remote bands within the energy cutoff
    $E_c=80\,\mathrm{meV}$.
    Purple, green, and orange curves denote system sizes
    $L=3$, $6$, and $9$, respectively.
    The projected model exhibits an approximately size-extensive structure
    factor throughout the displayed range of $u_0$, whereas the remote-band
    calculation shows that the KIVC is robust in a wide range of $u_0/u_1\lesssim 0.9$ and only after that its suppression is seen.}
    \label{fig:fig4}
\end{figure}

\newsect{Results}
Figure~\ref{fig:fig2} shows representative electron and trion spectra obtained from QMC and stochastic analytic continuation (SAC, see Ref.~\cite{huangAngle2025} for the technical details) at $u_0=80$, $90$, and $110~\mathrm{meV}$. For $u_0=80$ and $90~\mathrm{meV}$, the low-temperature ($T=0.5$ meV) electron spectra are fully gapped, while the trion spectral weight is strongly suppressed near $\Gamma$, consistent with the presence of KIVC correlations as shown in Fig.~\ref{fig:fig4} (b). At $T=3~\mathrm{meV}$, the KIVC correlations are weakened: since the KIVC breaks $U(1)$ symmetry, finite temperature will melt the long-range order, and as we have shown in previous work~\cite{shanTrion2026} that $T=3$ meV is sufficient to suppress the remaining power-law correlation. Under such conditions, both the electron and trion spectra develop low-energy Dirac-like features near $\Gamma$. In particular, the similar spectra at $u_0=80$ and $90~\mathrm{meV}$ demonstrate that the Dirac trion survives the inclusion of remote bands over an extended parameter range. This behavior changes qualitatively at the band-touching point $u_0=u_1=110~\mathrm{meV}$, where the flat--remote band separation $\gamma$ vanishes. The electron spectrum from the remote bands then retains substantial low-energy weight near $\Gamma$ at both temperatures, whereas the corresponding trion weight is strongly suppressed. Thus, near band touching, the dominant low-energy charged excitations acquire an itinerant remote-electron character.

To follow the evolution with increasing $u_0$ from the representative cases shown in Fig.~\ref{fig:fig2} toward the band-touching point $u_0=u_1=110~\mathrm{meV}$, Fig.~\ref{fig:fig3} shows the midpoint Green's function at $\Gamma$ for $u_0=60\text{--}110~\mathrm{meV}$, which is defined as 
\begin{equation}
-G(\Gamma,\beta/2)=\int d\omega\frac{A(\Gamma,\omega)}{2\cosh(\beta\omega/2)}\simeq\frac{\pi}{\beta}A(\Gamma,\omega=0).
\end{equation}
This quantity probes the thermally broadened low-energy spectral weight without stochastic analytic continuation. As shown in Fig.~\ref{fig:fig3}(a), at $T=0.5~\mathrm{meV}$ the remote-band electron contribution is negligible at small and intermediate $u_0$ but rises sharply upon approaching the band-touching limit. This demonstrates that only close to the band-touching limit, the remote-band electrons participate directly in the low-energy excitations and can no longer be neglected, marking the breakdown of the projected description. At $T=3~\mathrm{meV}$, the same evolution appears as a smoother crossover. 

As shown in Fig.~\ref{fig:fig3}(b), the projected and remote-band calculations yield comparable trion spectral weight over a broad range of $u_0$, demonstrating that remote bands do not qualitatively alter the trion physics in this regime. Near the band-touching limit, however, the trion contribution decreases while the remote-electron contribution increases. These opposite trends show that, only when the flat--remote band separation $\gamma \to 0$, do the dominant low-energy charged excitations switch from composite trions to itinerant remote-band electrons.

At charge neutrality, previous studies based on the projected model have identified the KIVC insulator as the ground state near the magic angle~\cite{bultinckGround2020,liaoCorrelation2021,liaoCorrelated2021,hofmann2022fermionic,panDynamical2022}. It is therefore important to determine whether the KIVC ground state remains stable when the remote bands are explicitly included. Figure~\ref{fig:fig4} compares the finite-size behavior of the KIVC structure factor in the projected and remote-band calculations at $T=0.5~\mathrm{meV}$. In the projected model [Fig.~\ref{fig:fig4}(a)], the structure factor increases with system size throughout the displayed range of $u_0$, consistent with persistent KIVC order. When the remote bands are retained [Fig.~\ref{fig:fig4}(b)], the KIVC correlations remain robust over a broad range of $u_0$, and are then suppressed at $u_0=100$ and $110~\mathrm{meV}$ and no longer increase with system size. These results indicate that, upon including the remote bands, the KIVC ground state remains stable over a wide range of $u_0/u_1 \lesssim 0.9$ but is lost at $u_0=100$ and $110~\mathrm{meV}$, implying a different ground state near band touching. 

We note that the ab-initio calculations and QTM experiment estimate \(u_0/u_1\) to lie in the range \(0.6\text{--}0.8\)~\cite{namLattice2017,koshinoMaximally2018,carrExact2019,xiao2026imaging}. This experimentally relevant range lies well within the regime where our results validate the flat-band projection, which captures both the KIVC ground state and the low-energy charged excitations, including the Dirac trion. A qualitative breakdown occurs only as \(u_0/u_1\) approaches 1: remote-band electrons acquire substantial near-zero-frequency spectral weight, while the KIVC ground state is absent. The remote-band degrees of freedom must therefore be explicitly retained near the band-touching limit, consistent with a heavy-fermion-like description.

\newsect{Discussion} 
By explicitly including remote BM bands in sign-problem-free momentum-space QMC, we establish the broad validity of flat-band projection in charge-neutral TBG. For \(u_0/u_1\lesssim0.9\), the projected calculations capture the low-energy electron and trion spectra, while the KIVC insulating state remains robust. Thus, a many-body gap comparable to the non-interacting flat–remote band separation does not, by itself, invalidate the projected description. In particular, the Dirac trion survives in the normal Mott state after intervalley-coherent order melts, demonstrating that this composite excitation is not an artifact of projection and supporting a mixed-valence description with active local charge fluctuations~\cite{zhao2025mixed,zhao2025ancillatheorytwistedbilayer,zhangEmergent2026}. Only near band touching does the explicit inclusion of itinerant remote-band electrons, as in the THF description, become necessary.

These results strengthen the motivation to explore composite charged excitations experimentally. Recent QTM measurements resolve interaction-renormalized electron spectra~\cite{inbarQuantum2023,xiao2026imaging}, including suppressed electron spectral weight upon doping certain integer-filling states, which motivates the possible role of carriers with weak electron overlap~\cite{shanTrion2026}. Although our calculations focus on charge neutrality, the robustness of Dirac trions against remote-band mixing provides a firmer basis for investigating this possibility at finite doping. QTM measurements on samples with controlled and independently characterized twist-angle and strain profiles~\cite{kapferProgramming2023,tran2024quantitative} would provide valuable comparisons with future calculations of the momentum-resolved spectra.

Our methodology and findings reconcile the different viewpoints on modeling interaction effects in TBG by connecting the projected and heavy-fermion limits within a unified quantum many-body framework. Our computational advance provides a foundation for future QMC studies away from neutrality, aiming at clarifying the nature of charge carriers and interpreting experimental spectra, and opens the way to studies of other correlated topological flat-band materials.

\newsect{Acknowledgment}
We thank Yahui Zhang, Zhi-Da Song, Yves H. Kwan and Francisco Guinea for the inspiring discussion. We thank Patrick Ledwith for prior collaborations on closely related work ~\cite{shanTrion2026}. S.B.S, C.H. and Z.Y.M. acknowledge the support from the Research Grants Council (RGC) of Hong Kong Special Administrative Region (SAR) of China (Project Nos. AoE/P701/20, C7037-22GF, 17302223, 17301924, 17301725, AoE/P-604/25-R, 17302026), the ANR/RGC Joint Research Scheme sponsored by RGC of Hong Kong and French National Research Agency (Project No. A\_HKU703/22) and the State Key Laboratory of Optical Quantum Materials at HKU. We thank HPC2021 system under the Information Technology Services at the University of Hong Kong~\cite{hpc2021}, as well as the Beijing Paratera Tech Corp., Ltd~\cite{paratera} for providing HPC resources that have contributed to the research results reported within this paper.

\nocite{Grenander1994,Roberts1998,cooleyAlgorithm1965}
\bibliography{bibtex}
\bibliographystyle{apsrev4-2}

\section{End Matter}

\subsection{Projected density operator and spectral functions}

Here we provide further details of the Hamiltonian and observables. For a momentum transfer $\bQ=\bq+\bG$, where $\bk$ and $\bq$ lie in the mBZ and $\bG$ is a moiré reciprocal lattice vector, the projected density fluctuation operator is
\begin{equation}
\delta\rho_{\bQ}
=
\sum_{\bk m n s\eta}
\lambda_{mn}^{s\eta}(\bk,\bQ)
\left(
c_{\bk ms\eta}^{\dagger}
c_{\bk+\bq,ns\eta}
-\frac{1}{2}\delta_{\bq,\mathbf{0}}\delta_{m,n}
\right).
\end{equation}
Here, $\bk+\bq$ is understood to be folded back into the mBZ. The form factor is $\lambda_{mn}^{s\eta}(\bk,\bQ)=\langle u_{ms\eta}(\bk)|u_{ns\eta}(\bk+\bQ)\rangle$, where $|u_{ns\eta}(\bk)\rangle$ is an eigenstate of the BM Hamiltonian and the second state is understood in the extended-zone convention. The band-resolved electron Green's function is defined as $G_n^{s\eta}(\bk,\tau)=-\langle T_\tau c_{\bk ns\eta}(\tau)c_{\bk ns\eta}^{\dagger}(0)\rangle$, while the trion Green's function is defined as $G_F^{\sigma s\eta}(\bk,\tau)=-\langle T_\tau F_{\bk\sigma s\eta}(\tau)F_{\bk\sigma s\eta}^{\dagger}(0)\rangle$. For both electron and trion Green's functions, the corresponding real-frequency spectral function satisfies~\cite{panDynamical2022}
\begin{equation}
G(\bk,\tau)
=
-\int_{-\infty}^{\infty}d\omega\,
\frac{\mathrm{e}^{-\omega\tau}}
{1+\mathrm{e}^{-\beta\omega}}
A(\bk,\omega).
\label{eq:GA}
\end{equation}
The spectral function $A(\bk,\omega)$ is obtained through stochastic analytic continuation (SAC), whose implementation is described in Ref.~\cite{huangAngle2025}.

\subsection{Energy-cutoff convergence}
\begin{figure}[H]
    \centering
    \includegraphics[width=\columnwidth]{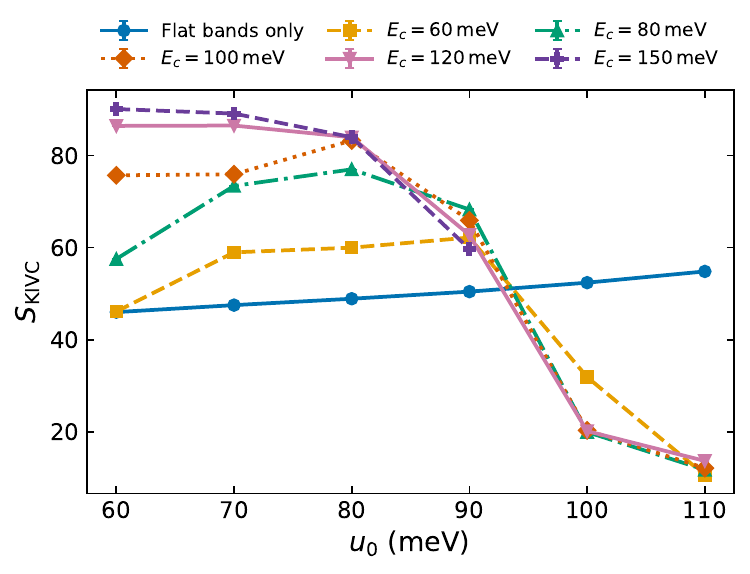}
    \caption{\textbf{Dependence of the KIVC correlation function on the band-energy cutoff.}
    The KIVC correlation function $S_{\mathrm{KIVC}}$ is shown as a function of $u_0$ at $L=6$, $T=0.5~\mathrm{meV}$, and $\theta=1.08^\circ$. The projected two-flat-band result is compared with calculations retaining all BM bands within $\lvert E\rvert\leq E_c$. The convergence trend with increasing $E_c$ and the persistent suppression of $S_{\mathrm{KIVC}}$ near the band-touching regime demonstrate that the qualitative behavior is insensitive to the band-energy cutoff.}
    \label{fig:cutoff-convergence}
\end{figure}

We examine the dependence of the KIVC correlations on the band-energy cutoff at fixed system size $L=6$. As shown in Fig.~\ref{fig:cutoff-convergence}, across the range $60\leq u_0\leq110~\mathrm{meV}$, $S_{\mathrm{KIVC}}$ exhibits a clear convergence trend with increasing $E_c$, while retaining its pronounced suppression at $u_0=100$ and $110~\mathrm{meV}$. The disappearance of the KIVC ground state near the band-touching regime is therefore not an artifact of the band-energy cutoff used in the main-text calculations.

\end{document}